# Hidden correlation between absorption peaks in achiral carbon nanotubes and nanoribbons

V. A. Saroka[1], A. L. Pushkarchuk[2], S. A. Kuten[1], and M. E. Portnoi[3]

[1]Institute for Nuclear Problems, Belarusian State University

Bobruiskaya 11, 220030, Minsk, Belarus, 40.ovasil@gmail.com

[2] Institute of Physical Organic Chemistry, National Academy of Sciences

Surganov Street 13, 220072 Minsk, Belarus

[3]School of Physics, University of Exeter

Stocker Road, EX4 4QL, Exeter, United Kingdom

**Abstract**

In this paper we study the effect of absorption peak correlation in finite length carbon nanotubes and graphene nanoribbons. It is shown, in the orthogonal π-orbital tight-binding model with the nearest neighbor approximation, that if the ribbon width is a half of the tube circumference the effect takes place for all achiral ribbons (zigzag, armchair and bearded), and corresponding tubes, starting from lengths of about 30 nm. This correlation should be useful in designing nanoribbon-based optoelectronics devices fully integrated into a single layer of graphene.

**Introduction**

Carbon nanotubes (CNTs) are promising nanostructures for post-silicon optolectronics [1–3]. However, although individual carbon nanotube devices show high performance, their large scale integration is still problematic [4]. For instance, due to carbon nanotubes cylindrical topology they cannot be incorporated into monolayer graphene. Although the cylindrical topology seems to be compatible with bilayer structures and, in principal, to admit tube formation by electron beam lithography of bilayer graphene [5], even in this case tube-integration cannot be achieved without introducing additional strain at the "tube-bilayer" junction and without the tubes becoming so-called collapsed tubes [6]. In contrast, graphene nanoribbons (GNRs) are more attractive structures for monolayer integration. Due to their plain geometry, they can be naturally embedded into a single layer of graphene without any lattice mismatch. For instance, they could be produced by the patterned adsorption of hydrogen [7–10] or by the reduction of fluorinated graphene with an electron beam [11]. Thus, mapping the properties of tubular structures onto planar structures could be beneficial. It has been shown in the literature that the electronic properties of both structures exhibit high degree of similarity. Namely, there is a perfect match between the energy band structures of the armchair graphene nanoribbons and zigzag carbon nanotubes [12]. Somewhat similar comparisons can also be found for zigzag graphene nanoribbons and armchair carbon nanotubes [13,14]. These studies, however, omit consideration of the optical properties. At the same time, the studies of the optical properties of ribbons mainly highlight the differences between tubes and ribbons: selection rules, the presence of edge to bulk state transitions, etc. [15–18]. A detailed theoretical analysis of the optical matrix elements revealed that the equivalence between armchair nanoribbons and zigzag tubes also extends to their optical properties, even if curvature and edge effects are taken into account [19]. For zigzag graphene nanoribbons, it has been recently reported that the optical absorption peaks, resulting from the bulk to bulk state transitions, correlate with those in armchair nanotubes when the quantized transverse momenta of their electrons are matched [20]. In the above-mentioned studies, infinitely long structures are considered while real carbon nanotubes and graphene nanoribbons have a large, but finite, length.

The purpose of the present paper is to investigate theoretically whether the aforementioned optical correlation takes place for finite length achiral structures. In the present analysis, additionally to zigzag and armchair edges, ribbons with so-called bearded edges [21] (or Klein edges [22]) are included. It is worth noting that such structures attract less attention in the literature since they are considered to be less stable in a free form when compared to zigzag and armchair ones. However, as predicted by first principles studies, ribbons with bearded edges (in addition to armchair and zigzag ribbons) should exist as fully-integrated into a single layer of graphene with patterned adsorption of hydrogen [10].

**Methods**

**Structural model**

The structure of an infinitely long carbon nanotube is described by the two chiral indices $(n, m)$ [23]. For finite length tubes, we supplement this conventional notation with an integer equal to the number of unit cells as defined for the infinite structure. Thus, we refer to a finite length tube cluster as $CNT(n, m, l)\ N$, where $n$ and $m$ are the standard chiral indexes, $l$ is the number of unit cells in the cluster, and $N$ is the total number of atoms in the cluster. The three cases of interest, corresponding to the achiral tubes, are presented in Fig. 1. (a), (b) and (c). As one can see from Fig. 1, the length (in nm) of a tube, $L$, is readily obtained by multiplying $l$ by the translation vector $T$ (as defined for the infinite structure). The total number of atoms in the structure $N = pl$, where $p$ is the number of atoms in the unit cell. In principle, $p$ can be expressed via the chiral indices $n$ and $m$, therefore the parameter $N$ is superfluous. However, it is useful to have it to hand for the estimation of cluster size and for computational purposes.

Graphene nanoribbons are usually characterized by their edge geometry and the width index, *w*, which counts the number of carbon atom pairs in the ribbon unit cell. Chiral, and more complicated, ribbons can also be constructed. However, there is no conventional notation for them yet; they could be described as in Refs. [24–30]. In what follows, we focus only on achiral ribbons with straight edges, and hence stick to the most common notation; we make slight amendments to incorporate the features of finite length structures. Thus, we refer to graphene nanoribbons with zigzag (Z), armchair (A) and bearded (B) edges (see Fig. 1 (d), (e) and (f)) as ZGNR$(w, l)$ $N$, AGNR$(w, l)$ $N$ and BGNR$(w, l)$ $N$, where *w* is the number of carbon atom pairs in the unit cell of the ribbon, $N$ and $l$ have the same meaning as for nanotubes. In this way, in fact, we define a graphene macromolecule similar to Refs. [13,14,31,32]. However, our notation is more general than that in Refs. [13,14,31,32] since it includes bearded edges and it counts atoms in pairs rather than quartets.

**Computational model**

The electronic band structures/energy levels and the electron wave functions of the nanostructures in question are calculated within the orthogonal $\pi$-orbital tight-binding model and the nearest-neighbor approximation. The hopping integral is set to $\gamma = 3.12$ eV [33]. In order to reveal the pure finite length effect for tubes and ribbons, we neglect the effects of intrinsic strain originating from the curvature in tubes [34] and edge relaxation in ribbons [35], i.e. all hopping integrals are the same throughout the structure. For the model study, we choose tubes with diameters $d > 0.6$ nm [see Fig. 1 (a), (b) and (c)] for which curvature effects, such as the mixing of $\sigma$- and $\pi$- orbitals, are insignificant [36]. In this paper, we solve the eigenproblem for the tight-binding matrix Hamiltonian numerically. For finite clusters the dipole moment matrix elements are calculated via the position operator matrix elements (see Ref. [37] for details) instead of the momentum (or velocity) matrix elements in the case of the infinitely long structures [20]. The absorption spectra for linearly polarized light (parallel to the longitudinal axis of the structure) are recovered at zero temperature ($T = 0$ K) by broadening the corresponding matrix elements by 0.02 eV to account for the effects of inhomogeneity, impurities and various scattering channels. The aforementioned approach is equally valid for optical absorption calculations in tubular and ribbon-like structures. Moreover, it provides a reasonable balance between the heavy first principles computations [38–41] and analytical treatments with a varying range of validity [18,20,42].

**Results and discussion**

The finite clusters depicted in Fig. 1 are based on infinitely long carbon nanotubes and graphene nanoribbons of the following types: (a) CNT(7,7), (b) CNT(12,0), (c) CNT(6,6), (d) ZGNR(6), (e) AGNR(11) and (f) BGNR(6). These infinitely long structures form "tube-ribbon" pairs, for which the number of atoms in the unit cell of the tube, $N_t$, is related to the number of atoms in the unit cell of the corresponding ribbon, $N_r$, as follows:

1) For CNT($w + 1, w + 1$) and ZGNR($w$), $N_t = 2N_r + 4$;

2) For CNT($w + 1, 0$) and AGNR($w$), $N_t = 2N_r + 4$ ;

3) For CNT($w, w$) and BGNR($w$), $N_t = 2N_r$.

Such relations between $N_t$ and $N_r$ result from the matching of the electron transverse momenta in each "tube-ribbon" pair. As one can see from Fig. 2, such matching leads to the energy bands of the infinitely long tubes and ribbons gaining much similarity. In particular, the band structures of armchair ribbons and zigzag nanotubes are almost identical [see Fig. 2 (b)]. Conversely, the band structures of zigzag and bearded ribbons are somewhat dissimilar to those of corresponding tubes. This can be attributed to the fact that the secular equations quantizing the electron transverse momenta in ZGNRs and BGNRs depend on the electron longitudinal momentum [20,43]. Nevertheless, it is clearly seen from Fig. 2 (a) and (c) that even for ZGNRs/BGNRs and corresponding armchair CNTs the energy bands follow the same pattern, so that their energy levels are approximately equal at the center of the Brillouin zone. The deviation from this pattern is observed close to the Fermi level for energy states that are the so-called edge states [22,44]. These states naturally do not have counterparts in the corresponding band structures of the tubes. It should be also mentioned that the tube energy bands with replicas in the ribbon band structure are double degenerate, while energy bands without such a replica are non-degenerate. The band structure matching presented in Fig. 2 (a) and (b) have been reported in Refs. [12] and [20] (see also Ref. [45]), respectively, while the matching in Fig. 2 (c) is presented for the first time. It should be also highlighted that for armchair CNTs and BGNRs, band structure matching requires a different relation between $N_t$ and $N_r$ compared to those of armchair CNTs and ZGNRs or zigzag CNTs and AGNRs.

Let us now proceed with the consideration of the optical properties of finite clusters based on the aforementioned tubes and ribbons. For such structures the unit cell element can be effectively introduced as shown in Fig. 1, therefore the same $N_t$ and $N_r$ as for infinitely long structures (with the same relations between them) can be used. In Fig. 3, we present the evolution of the absorption spectra of finite length tubes and ribbons with increasing length for the incident light linearly polarized parallel to the structure longitudinal axis. As one can see from Fig. 3, upon increasing the length of the cluster the absorption spectra converge to those of infinitely long structures, where the peak positions marked by the dashed vertical lines are given by the following analytical expressions:

$$\omega_{ZGNR,j} = 2\gamma\sin\left(\frac{\pi j}{w+1}\right) \qquad (1)$$

$$\omega_{AGNR,j} = 2\gamma \left|1 - 2\cos\left(\frac{\pi j}{w+1}\right)\right| \qquad (2)$$

$$\omega_{BGNR,j} = 2\gamma \sin\left(\frac{\pi j}{w}\right) \qquad (3)$$

where $\gamma = 3.12$ eV is the hopping integral [33] and $j = 1 \ldots w/2$ ($w$ is even) or $(w-1)/2$ ($w$ is odd). We derived these equations for infinitely long tubes and then adapted them for usage with the corresponding ribbons. As one can see, both second absorption peaks for armchair CNTs in Fig. 3 (a) and (c) (curves labeled with boxed numbers) do not have partners to form pairs, therefore $j = 1$ should be excluded when applying Eqs. (1) and (3) to ZGNRs and BGNRs, respectively. The intense peak at around 6.3 eV in Fig. 3 (b) originates from the flat bands seen in Fig. 2 (b) for CNT(12,0) and AGNR(11) at $E \approx \pm 3.1$ eV. We should note that, for the finite length AGNRs and zigzag CNTs, there are intense low-energy peaks in the far- (mid-)infrared frequency range ($\omega < 0.6$ eV). Similar peaks have been predicted for infinitely long zigzag CNTs and armchair ribbons. This is due to their being quasi-metallic as a result of intrinsic strain originating from the curvature effect in tubes and edge effect in ribbons [19]. Thus, we see that the effect of finite length gives rise to low-energy features in the absorption spectra, similar to those produced by intrinsic strain. The intense low-energy peak is also seen in Fig. 3 (a) and (c) for finite length armchair CNTs (curves labeled with boxed numbers). Such a peak is normally absent in the absorption spectra of the infinitely long armchair tubes [46]. However, a strong low-energy peak has been reported for infinitely long tubes subjected to an external magnetic field parallel to their longitudinal axes [47–50]. A somewhat similar peak can also be noticed in plots for finite length AGNRs of the metallic family, i.e. $w = 3p + 2$, where $p$ is an integer, with the third order nearest neighbors taken into account [51]. The origin of this peak in Ref. [51] is somewhat obscure, since it has been shown by Gunlycke et al. [52] that the third order nearest neighbor hopping integrals should open a narrow band gap in infinitely long metallic armchair ribbons which may cause such a peak. As follows from our results presented in Fig. 3 (b), the peak can arise from the pure finite length effect. Next we can notice that the low-energy peak in question is absent in the absorption spectra of ZGNR and BGNRs. This is because they possess edge states [see Fig. 2 (a) and (c)] and obey different optical selection rules. The selection rules for ZGNRs have been studied both numerically [15–17,53–55] and analytically [18,20,56]. As follows from our results in Fig. 3 (c), graphene nanoribbons with bearded edges obey the same selection rules as ZGNRs. Also, in such ribbons, the edge states do not contribute to the absorption of the parallel polarization of the incident light. This behaviour is evidence of the optical transition matrix elements in infinitely long BGNRs being zero at the centre of the Brillouin zone, like in armchair CNTs or ZGNRs [20]. Thus, both ZGNRs and BGNRs have selection rules that are different from those in armchair CNTs. Moreover, as one can see from Fig. 2 (a) and (c), the band structures (energy levels for finite clusters) are not the same. Obviously, the band structures matching shown in Fig. 2 (b) is much better. Yet, in the absorption spectra for armchair tubes and zigzag/bearded ribbons we have correlated peaks (marked with vertical dashed lines) originating from the bulk-to-bulk state transitions, i.e. not involving edge states. Thence, the correlation of the absorptions peak positions in ZGNRs/BGNRs and armchair CNTs is a non-trivial effect, which can be dubbed as a hidden correlation.

Let us now focus on the evolution of the absorption spectra in question with the cluster lengths. For the achiral structures shown in Fig. 3, one can unambiguously recognize the absorption spectra of the infinitely long structure, starting from curves No. 3, boxed for tubes and circled for ribbons. These curves correspond to $L = 15.9$ nm for BGNR, ZGNR and armchair CNTs and to $L = 27.7$ nm for AGNRs and zigzag CNTs. Hence, the correlated peaks close to those given by the infinitely long structure model can be observed starting from lengths of about 30 nm. Since many of the experimental studies dealing with short tubes or ribbons operate with lengths > 100 nm [57,58], one can conclude that the correlated peak positions in achiral tubes and ribbons may be observed in currently available samples if the above-mentioned relations for $N_t$ and $N_r$ could be satisfied. As seen from Fig. 3 (b), in the case of finite length armchair ribbons and zigzag tubes the correlated peaks can be observed even below the length threshold of 30 nm. The number of correlated peaks in such structures, however, should be larger than that given by Eq. (2).

The correlation reported here between the absorption peak positions in achiral carbon nanotubes and graphene nanoribbons implies that an achiral carbon nanotube can be decomposed into two corresponding nanoribbons with optical properties very similar to those of the initial tube. Such decompositions are explicitly presented in Fig. 4 for all types of structure discussed above. It is readily seen from Fig. 4 that $N_t = 2N_r + 4$ (for ZGNR($w$) and CNT($w + 1, w + 1$); AGNR($w$) and CNT($w + 1,0$)) and $N_t = 2N_r$ (for BGNR($w$) and CNT($w, w$)) relations between the number of atoms in the tube and ribbon unit cells correspond to the width of the ribbon, $W$, being half of the circumference of the tube, $C$. This decomposition should have practical implications. As has been mentioned in the introduction, the cylindrical topology of the nanotubes does not allow their smooth integration into a graphene monolayer. In contrast, the planar structure of the ribbons is naturally suitable for such an embedment. In other words, the correlation between absorption peaks revealed here allows one to map a cylindrical tube onto a planar ribbon, while preserving some of its optical properties. This property should be useful for the large scale integration of ribbon-based optoelectronic devices into a flexible graphene substrate with the unique possibility to mimic the performance of tube-based devices.

A correlation, similar to that reported here, might also take place for carbon nanotubes and superlattices based on graphene nanoribbons [27–30] or tubes and ribbons based on other 2D materials [59]. For instance, many features of the graphene nanoribbons and quantum dots have also been revealed in phosphorene analogues [60–63]. Moreover, the stability of phosphorene nanotubes has been recently predicted by first principles calculations [64].

Finally, we should notice that more sophisticated models can be applied in the future to study the relations between tubular and planar structures. However, the most advanced models, providing the best fit to the experiments by taking into account many-body and excitonic effects, are limited to a few hundreds of atoms in the clusters. In order to reveal the finite length effect, we had to deal with a significantly greater number of atoms [see Fig. 3 caption]. Our results imply that to simplify such calculations, while testing the peak correlation in question, the structures can be treated as infinitely long and the periodic boundary conditions can be used for their longitudinal direction (Bloch theorem). This should also allow the treatment of structures which are large in the transverse direction, thereby increasing the number of correlated peaks and increasing the probability of identifying the correlation between the absorption peaks in tubes and ribbons. The present research verifies that this effect is not an accidental feature; it also reveals inherent trends for the peak correlation effect thereby providing an important reference point for the next level of studies.

**Conclusions**

In summary, we have shown that the decomposition of all achiral carbon nanotubes onto two achiral graphene nanoribbons of equal width [CNT $(w + 1,0) \rightarrow$ AGNR $(w)$, CNT $(w + 1, w + 1) \rightarrow$ ZGNR$(w)$/BGNR$(w)$] maintains the positions of the absorption peaks, giving rise to correlation between the absorption peak positions in thus related tubes and ribbons. In the tight-binding model, with the nearest neighbor approximation, this correlation is also present for realistic finite length tubes and ribbons ($L > 30$ nm) and therefore should be useful in the spectroscopic characterization of such structures. For instance, Kataura plots [65], if properly rescaled, may be applied to the characterization of achiral nanoribbon width. The revealed correlation also implies that planar structures, fully integrated into a graphene monolayer, can mimic the optical response of tubular structures. We point out that the correlation effect may be tested on isolated tubes and ribbons. Recently, graphene ribbons with zigzag and armchair edge topology have been produced by self-assembling techniques with atomic precision [66,67]. At the same time, the carbon nanotube samples have also been drastically improved in quality. In addition to the gradual advancement of carbon nanotube sorting techniques [68], the first successful attempt at their synthesis with pre-defined chirality has been reported [69].

## Acknowledgments

The authors are grateful to C. A. Downing for a number of fruitful discussions on this subject and R. Keens for proofreading the manuscript. We also thank P. P. Kuzhir and K. G. Batrakov for stimulating discussions and M. V. Shuba for constructive feedback.

## Declaration of interest statement

This work was supported by the EU FP7 ITN NOTEDEV (Grant No. FP7-607521), EU H2020 RISE project CoExAN (Grant No. H2020-644076), FP7 IRSES projects CANTOR (Grant No. FP7-612285), QOCaN (Grant No. FP7-316432), InterNoM (Grant No. FP7-612624); Graphene Flagship (Grant No. 604391) and partially by the Belarus state program of scientific investigations "Convergence-2020".

**Figures**

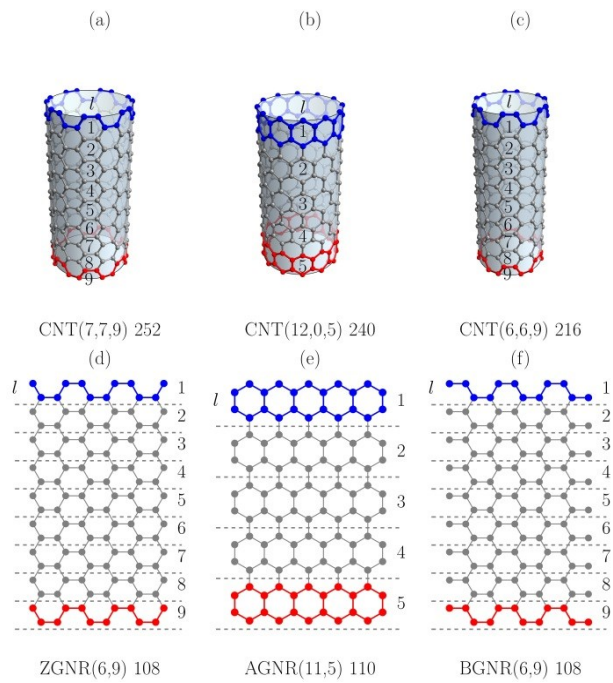

Fig. 1:

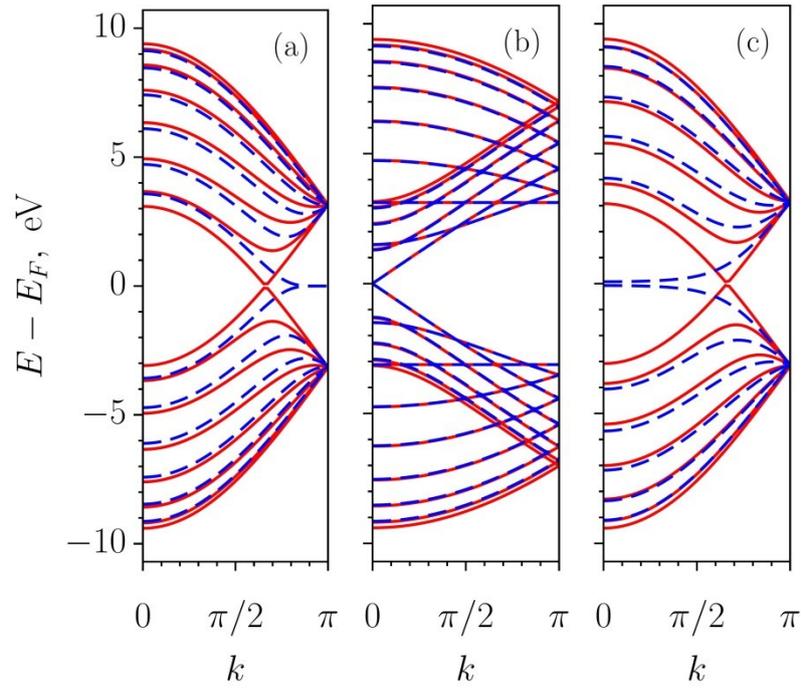

Fig. 2:

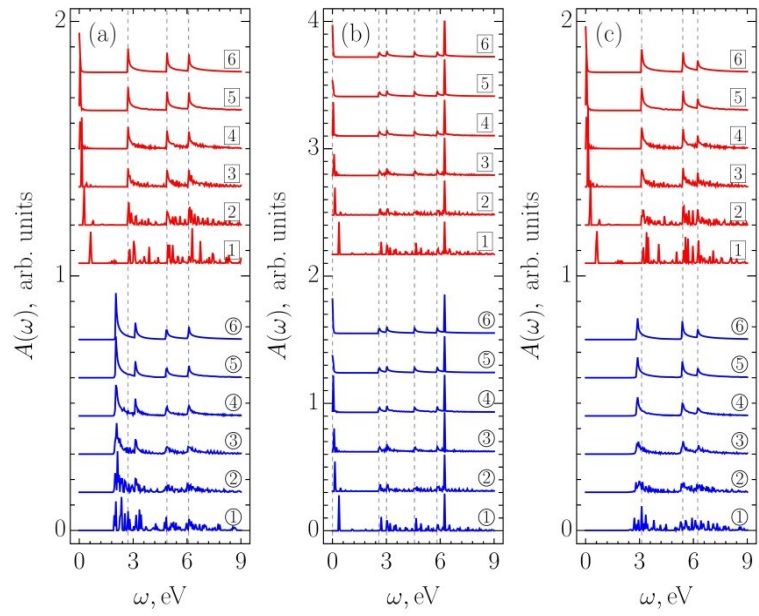

Fig. 3:

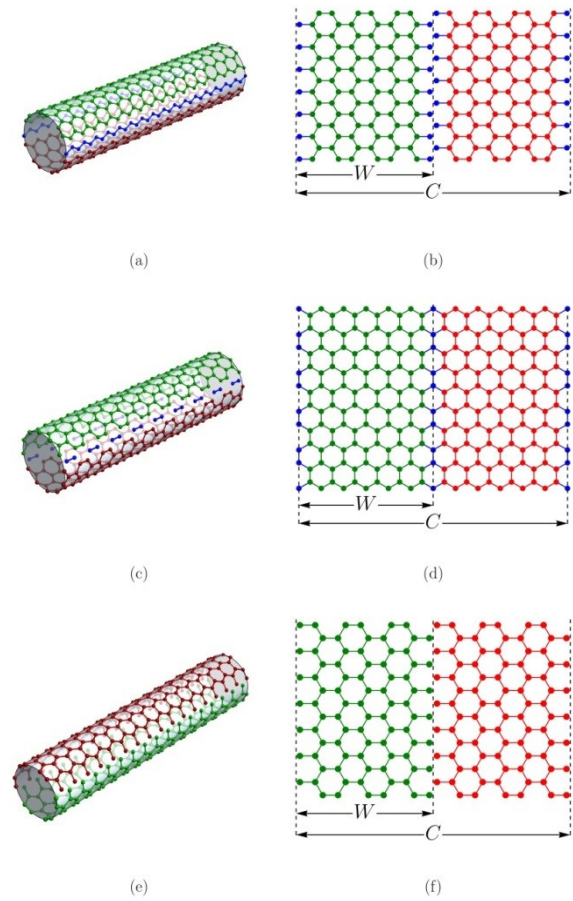

(a)  (b)

(c)  (d)

(e)  (f)

Fig. 4:

**Figure captions (as a list)**

1. Fig. 1: The atomic structures of the clusters in question: (a), (b) and (c) finite tubes and (d), (e) and (f) ribbons. The red and blue colors highlight the unit cells of the corresponding infinitely long structures.

2. Fig. 2: The band structures of infinitely long tubes (solid red curves) and ribbons (dashed blue curves) with the matched electron transverse momenta: (a) CNT(7,7) and ZGNR(6); (b) CNT(12,0) and AGNR(11); (c) CNT(6,6) and BGNR(6).

3. Fig. 3: The absorption spectra of finite length achiral carbon nanotubes (red, boxed) and graphene nanoribbons (blue, circled): (a) CNT(7,7, $l$) $N$ and ZGNR(6, $l$) $N$; (b) CNT(12,0, $l$) $N$ and AGNR(11, $l$) $N$ (c) CNT(6,6, $l$) $N$ and BGNR(6, $l$) , where

$l = 13, 33, 65, 129, 257, 513$ for boxed and circled labels 1,2,3,4,5, and 6, respectively. In ascending order of the labels the total number of atoms in the clusters are $N = 156, 396, 780, 1548, 3084, 6156$ for $ZGNR(6, l)$ and $BGNR(6, l)$;

$N = 286, 726, 1430, 2838, 5654, 11286$ for $AGNR(11, l)$;

$N = 364, 924, 1820, 3612, 7196, 14364$ for $CNT(7,7, l)$;

$N = 624, 1584, 3120, 6192, 12336, 24624$ for $CNT(12,0, l)$;

$N = 312, 792, 1560, 3096, 6168, 12312$ for $CNT(6,6, l)$;

4. Fig. 4: The decomposition of achiral CNTs into two GNRs: (a),(b) CNT(7,7) into ZGNRs(6); (c),(d) CNT(12,0) into AGNRs(11); (e),(f) CNT(6,6) into BGNRs(6). The 3D and 2D models are presented in panels (a), (c), (e) and (b), (d), (f), respectively. The blue color highlights atoms to be removed from the structures.